\documentclass[twocolumn,pra,aps,superscriptaddress,showpacs]{revtex4-1}

\usepackage{mathptmx}
\usepackage{subfigure}
\usepackage{psfrag,graphicx}
\usepackage{dcolumn}
\usepackage{amsmath,amssymb}
\usepackage{bm}
\usepackage{color}
\usepackage{latexsym}
\usepackage{epstopdf}
\usepackage{color}
\usepackage[english]{babel}
\usepackage{latexsym}
\usepackage{psfrag,graphicx}
\usepackage{subfigure}
\usepackage{amsmath}
\usepackage{amssymb}
\usepackage{amsfonts}
\usepackage{bm}
\usepackage{natbib}
\usepackage{epstopdf}
\usepackage{appendix}

\definecolor{mygrey}{gray}{0.35}
\definecolor{myblue}{rgb}{0.2,0.2,0.8}
\definecolor{myzard}{cmyk}{0,0,0.05,0}
\definecolor{mywhite}{rgb}{1,1,1}
\definecolor{mywhite}{rgb}{1,1,1}
\definecolor{myred}{rgb}{1,0.,0.3}

\usepackage[colorlinks=true,citecolor=myblue,linkcolor=myred]{hyperref}

\def\be{\begin{equation}}
\def\ee{\end{equation}}
\def\ba{\begin{align}}
\def\enda{\end{align}}
\def\bi{\begin{itemize}}
\def\ei{\end{itemize}}

\begin{document}


\title{Adiabatic quantum metrology with strongly correlated quantum optical systems}
\author{P. A. Ivanov}
  \affiliation{
  Department of Physics, Sofia University, James
Bourchier 5 Boulevard, 1164 Sofia, Bulgaria}

\author{D. Porras}
\affiliation{Departamento de F\'isica Te\'orica I,
  Universidad Complutense,
  28040 Madrid,
  Spain}


\begin{abstract}
We show that the quasi-adiabatic evolution of a system governed by the Dicke Hamiltonian can be described in terms of a self-induced quantum many-body metrological protocol.
This effect relies on the sensitivity of the ground state to a small symmetry-breaking perturbation at the quantum phase transition, that leads to the collapse of the wavefunciton into one of two possible ground states.
The scaling of the final state properties with the number of atoms and with the intensity of the symmetry breaking field, can be interpreted in terms of the precession time of an effective quantum metrological protocol.
We show that our ideas can be tested with spin-phonon interactions in trapped ion setups.
Our work points to a classification of quantum phase transitions in terms of the capability of many-body quantum systems for parameter estimation.
\end{abstract}

\maketitle

\section{Introduction}
Experimental progress in the last years has provided us with setups in Atomic, Molecular an Optical physics in which interactions
between many particles can be controlled and quantum states can be accurately initialized and measured.
Those experimental systems have an exciting outlook for the analogical quantum simulation of many-body models \cite{Cirac12natphys,SchneiderRPP,Blatt12natphys}.
For example trapped ion setups can be used to simulate the physics of quantum magnetism
\cite{Porras04aprl,Friedenauer08natphys,Islam11natcom,Britton12nat} and quantum structural phase transitions \cite{Porras12bprl,Bermudez12prl,Ivanov2013} by means of spin-dependent forces.
A more established practical application of atomic systems is in precision measurements for atomic clocks and frequency standards.
The effect of quantum correlations on the accuracy of interferometric experiments has been investigated in the field of quantum metrology
\cite{Giovannetti_2011,Giovannetti_prl2006}.
Here, entangled states may yield a favorable scaling in the precision of a frequency measurement compared to uncorrelated states
\cite{Wineland94pra,Bollinger_96,Leibfried_2004,Roos06}.
In view of this perspective a question arises, namely,
whether we can find applications of strongly correlated states of quantum simulators
for applications in quantum metrology.

A natural direction to be explored is the use of quantum phase transitions \cite{sachdev.book} in atomic systems.
Intuition suggests that close to a phase transition a system becomes very sensitive to small perturbations.
In particular, if there is a phase transition to a phase with spontaneous symmetry breaking, we may expect that
any tiny perturbation leads the system to collapse to one of several possible ground states.
Actually, quantum states typically considered for quantum metrology, such as NOON states, have a close relation to ferromagnetic phases of mesoscopic Ising models.
However, frequency measurements typically rely on dynamical processes, for example in Ramsey spectroscopy \cite{Wineland94pra,Peik06jphysb}. Thus the conditions
under which an atomic system remains close to the ground state of a many-body Hamiltonian must be carefully studied in view of possible metrological applications.

In this work we present a proposal to fully exploit the spontaneous symmetry breaking of a discrete $\mathbb{Z}_2$ symmetry to implement a
quantum metrology protocol with a system described by the Dicke Hamiltonian \cite{Dicke,Garraway2011}. The latter is the simplest model
showing a quantum phase transition, and remarkably it can be implemented in a variety of experimental setups in atomic physics,
from trapped ions \cite{Ivanov2013} to ultracold atoms
\cite{Baumann_2010, Baumann_2011}. Our scheme relies on an
adiabatic evolution which takes the system across a quantum phase transition where $\mathbb{Z}_2$ is spontaneously broken.
We show that the system is very sensitive to the presence of a symmetry breaking field, $\delta$,
such that it self-induces a many-body Ramsey spectroscopy protocol which can be read out at the end of the process.
Within the adiabatic approximation, we show that the ground state multiplet of the Dicke model can be approximated by an effective
two-level system, something that allows us to obtain an analytical result for the measured signal as a function of the number of atoms $N$.

Our proposal can work in two different ways:
{\it (i) Quasi-adiabatic method.-}
Non-adiabatic effects within the two-level ground state multiplet lead to variations in the final magnetization. By reading out the final state we recover the scalings corresponding to the Heisenberg limit of parameter estimation.
{\it
(ii) Full adiabatic method.-} Here we consider the information that is obtained by a single-shot measurement of the final magnetization. The system collapses into one of the possible symmetry broken states, and this allows to get the sign of the symmetry breaking field within a measurement time that scales inversely proportional to the number of particles, $t_m \propto 1/N$.

This article is structured as follows.
In section \ref{DMQS} we introduce the Dicke Hamiltonian and the concept of spontaneous symmetry breaking.
In section \ref{TL} we discuss the low energy spectrum of the normal and superradiant phases of the Dicke model, and show that close to the adiabatic limit, the dynamics of the system can be described by a two-level approximation.
In section \ref{PMS} we show how the evolution of the gap in the Dicke Hamiltonian allows us to think of a quasi-adiabatic evolution separated in a (fast) preparation stage followed by a (slow) measurement stage in which the system is sensitive to a small perturbation. In section \ref{QMP} we discuss the two different schemes for getting information of the symmetry breaking field that can be envisioned from the low-energy physics of the Dicke Hamiltonian.
\begin{figure}[h]
\includegraphics[width=0.45\textwidth]{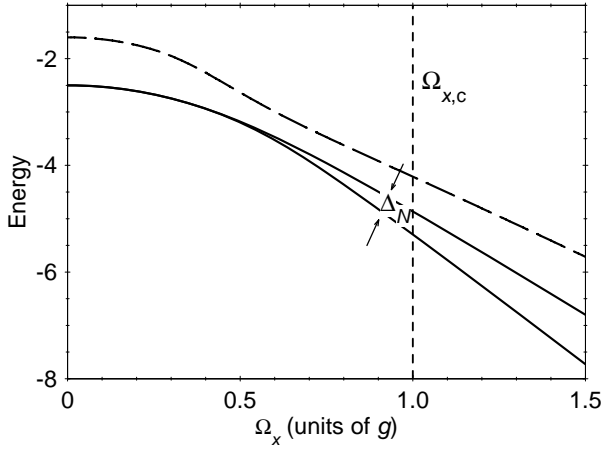}
\caption{a) The low energy spectrum (in units of $g$) of the single mode Dicke model for $\omega=4g$ and $N=10$ atoms as a function of $\Omega_{x}$. The two lowest lying states are separated by energy splitting $\Delta_{N}$. In a superradiant phase $\Omega_{x}<\Omega_{x,{\rm c}}$ the gap $\Delta_{N}$ can be calculated analytically by $N$-order perturbation theory. The dashed line is the third excited energy level.}
\label{fig1}
\end{figure}
Section \ref{PITI} is devoted to the outline of a physical implementation of our ideas with trapped ion crystals interacting with lasers that induce spin-phonon couplings.
Finally, in section \ref{Conclusion} we summarize
our results and present an outlook for further directions of the ideas presented here.
\section{Dicke model for quantum spectroscopy}\label{DMQS}
We start by reviewing the celebrated Dicke Hamiltonian describing an ensemble of $N$ two-level atoms coupled
to a single bosonic mode ($\hbar=1$ from now on),
\begin{eqnarray}
&&H=H_{\rm D} + H_{\delta},\nonumber \\
&&H_{\rm D}=\omega a^{\dag}a+\Omega_{x}J_{x}+\frac{2g}{\sqrt{N}}(a^{\dag}+a)J_{z},\nonumber \\
&&H_{\delta}=\delta J_{z}. \nonumber \\
\label{H}
\end{eqnarray}
$H_{\rm D}$ is the Dicke Hamiltonian, whereas $H_\delta$ is an additional symmetry breaking perturbation.
$a^{\dag}$ and $a$ are creation and annihilation operators corresponding to an oscillator with frequency $\omega$.
Collective spin operators
$\vec{J} = (J_{x},J_{y},J_{z})$ are defined by
\begin{equation}
J_{\beta} = \frac{1}{2}\sum_{i = 1}^{N} \sigma_{i}^{\beta},
\label{def.J}
\end{equation}
\begin{figure}[h]
\includegraphics[width=0.50\textwidth]{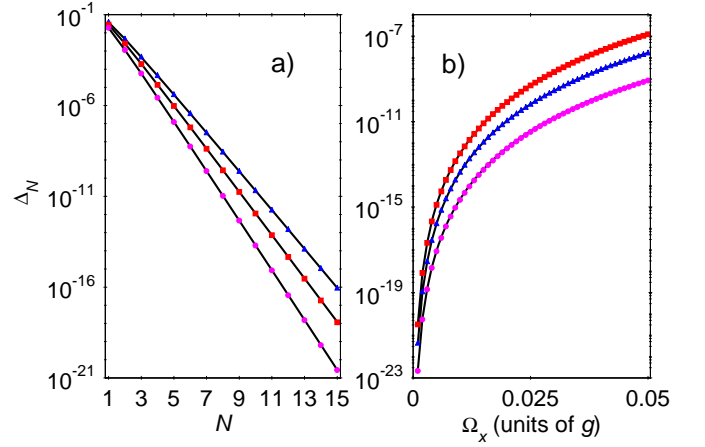}
\caption{a) The energy gap $\Delta_N(g,\omega,\Omega_x)$ of the Dicke Hamiltonian as a function of the number of atoms $N$
with $\omega=6g$ for various $\Omega_{x}$. The numerical results for $\Omega_x=0.02g$ (circles),
$\Omega_x=0.03g$ (square), $\Omega_x=0.04g$ (triangle) are compared with the analytical solution (\ref{gap.1}) (solid lines). b)
Numerical result for $\Delta_N(g,\omega,\Omega_x)$
as a function of $\Omega_{x}$ for $N=8$ atoms and $\omega=4g$ (black circles), $\omega=6g$ (blue triangles), $\omega=8g$ (red squares)
compared with (\ref{gap.1}) (solid lines). Values for $\Delta_N$ are given in units of $g$.}
\label{fig2}
\end{figure}where $\sigma_{i}^{\beta}$ ($\beta=x,y,z$) is the Pauli operator for each atom $i$.
$g$ and $\Omega_{x}$ are the intensive spin-boson coupling and transverse field, respectively.
The term $H_{\delta}$ describes the coupling to a longitudinal field  $\delta$, where we assume
the latter to be small in a sense to be precisely defined below.

The Dicke Hamiltonian is the simplest many-particle model with a discrete $\mathbb{Z}_2$ symmetry.
The latter is implemented by the parity operator defined by
\begin{eqnarray}
&&\Pi = \Pi_{\rm s} \otimes \Pi_{\rm b} , \nonumber \\
&&\Pi_{\rm s}  = \sigma_{1}^{x} \otimes \dots \otimes \sigma_{N}^{x}, \ \
  \Pi_{\rm b} = (-1)^{a^\dagger a}.
\label{parity.operator}
\end{eqnarray}
Since $\Pi H_{\rm D} \Pi = H_{\rm D}$, parity is a good quantum number. The discrete $\mathbb{Z}_2$ symmetry plays a decisive role in the discussion below.

In the limit $N \rightarrow \infty$ the mean-field solution becomes exact \cite{Hepp_73,Emary_2003}.
In this work we consider the evolution of the system with fixed $g$, $\omega$, and varying values of the transverse field, $\Omega_x$. In this case mean-field theory predicts a quantum phase transition at the critical point
$\Omega_{x,{\rm c}} = 4 g^2/\omega$. The latter separates a normal, or weak coupling phase ($\Omega_x \gg \Omega_{x, {\rm c}}$) with
$\langle J_z \rangle, \langle a \rangle = 0$, from the superradiant, or strong coupling phase, ($\Omega_x \ll \Omega_{x, {\rm c}}$) with
 $\langle J_z \rangle, \langle a \rangle \neq 0$.

Since $\Pi a \Pi = -a$, $\Pi J_z \Pi = -J_z$, the mean-field solution breaks the parity symmetry. This effect can be understood in the following way. Consider $|\Psi_\delta\rangle$, the ground state of the Hamiltonian (\ref{H}) with a finite longitudinal field $\delta$.
In the superradiant phase ($\Omega_x < \Omega_{x, {\rm c}}$) the following limit holds,
\begin{equation}
\lim_{\delta \to 0} \lim_{N \to \infty}
\langle \Psi_\delta | a | \Psi_\delta \rangle \neq 0,
\label{SSB}
\end{equation}
which implies that in the thermodynamical limit, an infinitesimal perturbation breaks the parity symmetry. Below we give an explicit proof of this result, which however is implicit in the fact that mean-field theory becomes exact as $N \to \infty$.

\section{Low-energy spectrum of the Dicke model}\label{TL}
\begin{figure}[h]
\includegraphics[width=0.50\textwidth]{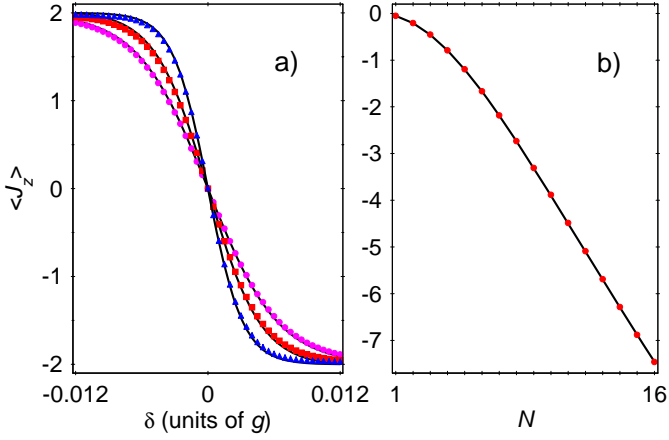}
\caption{a) The mean-value of $J_z$ as a function of $\delta$ for $\omega =3g$, $\Omega_{x}(0)=9g$ and $N=4$ for various $\gamma$.
We compare the numerical solution of the time-dependent Schr\"{o}dinger equation with the Hamiltonian (\ref{H}) for $\gamma=0.02g$
(triangle), $\gamma=0.03g$ (square) and $\gamma=0.04g$ (circle)
with the solution of the two-state problem Eq. (\ref{solution}) (solid lines).
b) The scaling of the measured signal as a function of $N$, for $\gamma=0.03g$, $\omega=10g$ and $\delta=2.0\times10^{-3}g$.
The red circles represent the numerical result, while the solid curve is the analytical solution, Eq. (\ref{Jz}). }
\label{fig3}
\end{figure}
In this section we show an effective description of the adiabatic quantum dynamics of $H_{\rm D} + H_\delta$ in terms of an effective two-level system.

First, we note that $H_{\rm D}$ commutes with the total angular momentum operator $\vec{J}^2$.
Let us consider the eigenstates of
$H_{\rm D} + H_\delta$ in the basis
$\{ | j, m \rangle | n \rangle_{\rm b} \}$, where $|j,m\rangle$ are the eigenstates of $\vec{J}^2$, $J_z$, and $| n \rangle_{\rm b}$ are the Fock states of the harmonic oscillator.
We will study the evolution of the system starting with a fully polarized state with $j = N/2$, such that conservation of $\vec{J}^2$ ensures that we remain within the $j= N/2$ subspace. The dimension of the spin Hilbert space is thus
$N + 1$, and the system is amenable to be studied with numerical diagonalization.

In the following we study the low-energy spectrum of $H_{\rm D}$ as a function of $\Omega_x$, something that will allow us to get an effective description of the full Hamiltonian $H = H_{\rm D} + H_\delta$ in the superradiant phase.
We define the two lowest eigenstates of
$H_{\rm D}$, $| \Psi_{{\rm gs},\mp}  \rangle$, with energies
$E_{{\rm gs},\mp}$.
The energy gap is
$\Delta_N(g,\omega,\Omega_x) = E_{{\rm gs},+}-E_{{\rm gs},-}$, for clarity in the calculations below we write it explicitly as a function of the parameters in the Dicke Hamiltonian.
\begin{figure}[h]
\includegraphics[width=0.45\textwidth]{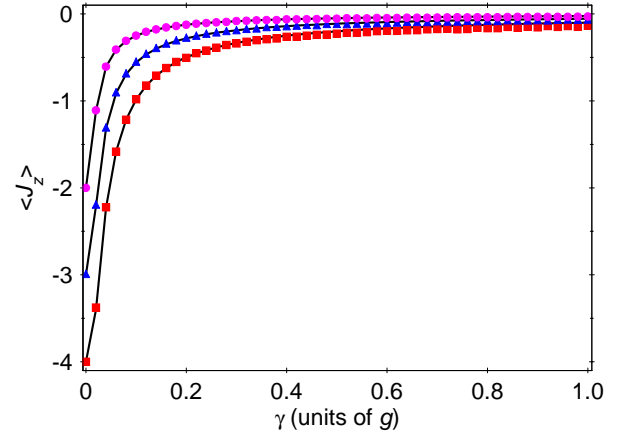}
\caption{The expectation value of $J_{z}$ as a function of $\gamma$ for $\omega =10g$, $\Omega_{x}(0)=9g$, $\delta=2.0\times10^{-3}g$ and various $N$. The exact solution for $N=4$ (circle), $N=6$ (triangle) and $N=8$ (square) is compared with
Eq. (\ref{solution}) (solid lines).}
\label{fig4}
\end{figure}

\subsection{Non-interacting limit ($\Omega_x \gg \Omega_{x,{\rm c}}$)}
We study first the limit $\Omega_x \to \infty$, or alternatively $g=0$.
Assuming $\Omega_x > 0$, the lowest energy state is the fully polarized spin-state in the $x$ direction,
\begin{equation}
| \Psi_{{\rm gs},-}^{\rm [g = 0]} \rangle
= \bigotimes_{k} | - \rangle_{x,k} | 0 \rangle_{\rm b} ,
\label{psi0.1}
\end{equation}
where $\left| \pm \right\rangle_{x,k}$ are the eigenstates of $\sigma_{k}^{x}$ and
$| n \rangle_{\rm b}$ is the Fock state of the bosonic
mode with occupation $n$.
The second lowest energy state is either a spin-wave if $\Omega_x < \omega$,
\begin{equation}
| \Psi_{{\rm gs},+} ^{\rm [g = 0]} \rangle =
\frac{1}{\sqrt{N}}
\sum_k | - \rangle_{x,1} \dots | + \rangle_{x,k} \dots | - \rangle_{x,N} | 0 \rangle_{\rm b},
\end{equation}
or an excitation of the harmonic oscillator if $\omega < \Omega_x$,
\begin{equation}
| \Psi_{{\rm gs},+} ^{\rm [g = 0]} \rangle =
\bigotimes_{k} | - \rangle_{x,k} | 1 \rangle_{\rm b} ,
\end{equation}
being the gap $\Delta_N(0,\omega,\Omega_x) = \Omega_x$, or $\omega$, respectively. In any case the two lowest energy states have opposite parity.

\subsection{Strong-interacting limit ($\Omega_{x,{\rm c}} \gg \Omega_x$)}
We discuss in more detail the superradiant phase, which is the most relevant for our quantum metrology protocol.

Consider first the limit $\Omega_x = 0$. Within the $j = N/2$ subspace, the spectrum of $H_{\rm D}$ corresponds to a set of Fock states of the harmonic oscillator, displaced by an amplitude proportional to the quantum number $m = -N/2, \dots, N/2$,
\begin{equation}
| \Phi_{n,m} \rangle = D\left( -\frac{2 g}{\omega \sqrt{N}} m \right)
|\frac{N}{2}, m \rangle | n \rangle_{\rm b},
\end{equation}
where we have defined the displacement operator
$D(\alpha) = e^{\alpha a^\dagger-\alpha^* a }$.
The eigenenergies are
\begin{equation}
E_{n,m} = n  \ \omega - \frac{g^2 N}{\omega} \left( \frac{m}{N/2} \right)^2 .
\end{equation}
We find two degenerate ground states, corresponding to
$m = \pm N/2$,
\begin{equation}
|\Psi_{{\rm gs},\pm}^{\rm [\Omega_x = 0]}\rangle =
D \left( \mp \sqrt{N} g / \omega \right)
\vert \frac{N}{2} , \pm \frac{N}{2} \rangle \vert 0 \rangle_{\rm b},
\label{ground.state.1}
\end{equation}
with energies $E_{{\rm gs},\pm} = -(g^2/\omega) N$.

Let us consider now the effect of a small transverse field, $\Omega_x$,
in the low-energy spectrum. We expect the energy gap, $\Delta_N(g,\omega,\Omega_x)$, to be lifted by the coupling of the two degenerate states by the term $\Omega_x J_x$ in $H_{\rm D}$.
However, note that the operator $J_x$ has to flip all spins to bring
$|\Psi_{{\rm gs},+}^{[\Omega_x = 0]} \rangle$ to
$| \Psi_{{\rm gs},-}^{[\Omega_x = 0]}\rangle$, such that
\begin{equation}
\langle \Psi_{{\rm gs},+}^{\rm [\Omega_x = 0]}| J_x^M | \Psi_{{\rm gs},-}^{\rm [\Omega_x = 0]} \rangle = 0,
\end{equation}
if $M \leq N-1$.
The first nonzero contribution is thus of order $N$.
$N$th order perturbation theory allows us to estimate the following scaling (see Appendix \ref{EG})
\begin{equation}
\Delta_N(g,\omega,\Omega_x) = f_N \left( g/\omega \right)
\Omega_x \left( \frac{\Omega_x}{\Omega_{x,c}} \right)^{N-1},
\label{gap.1}
\end{equation}
where $f_N(g/\omega)$ is a scaling function that describes the dependence of the gap on the ratio $g/\omega$. An explicit expression for $f_N(g/\omega)$ can be found in the particular case $g \ll \omega$,
\begin{equation}
f_N(g/\omega) = 2 e^{-2 \left( \frac{g}{\omega}\right)^2}
\frac{N^{N+1}}{2^{N} N!}.
\end{equation}
The latter corresponds to the limit in which the harmonic oscillator can be adiabatically eliminated, such that $H_{\rm D}$ is equivalent to an infinite range Ising spin Hamiltonian. For other values of $g/\omega$, one can use numerical calculations to estimate the exact form of the energy gap.
Note that the effect of a finite gap, $\Delta_N$, is to restore the parity symmetry by creating ground states that are linear combinations of
$| \Psi_{{\rm gs},-}^{[\Omega_x = 0]} \rangle$,
$|\Psi_{{\rm gs},+}^{[\Omega_x = 0]} \rangle$.

The most important feature of the superradiant phase, is thus the vanishing of the gap in the thermodynamical limit, analogously to the situation found, for example, in the short-range quantum Ising model \cite{sachdev.book}.
In a finite size system, $\Delta_N(g,\omega,\Omega_x)$ monotonically decreases as we decrease $\Omega_x$ from the value $\Omega_{x,{\rm c}}$.
The monotonic behavior of the gap with respect to the transverse field is actually valid along the whole phase diagram, and not only within the superradiant phase. This is shown in Fig. \ref{fig1}, where we present the evolution of the low-energy spectrum of $H_{\rm D}$.

%
%
%

Within the superradiant phase we can thus project the Hamiltonian $H = H_{\rm D} + H_{\delta}$ into the ground state multiplet to get the effective Hamiltonian,
\begin{equation}
H_{\rm eff} =
\frac{\Delta_N(g,\omega,\Omega_x)}{2}\sigma^x + \frac{N\delta}{2} \sigma^z,
\label{Heff}
\end{equation}
where Pauli operators act over the Hilbert subspace
$\{
|-\rangle, |+ \rangle \}$ =
$\{| \Psi_{{\rm gs},-}^{[\Omega_x = 0]} \rangle,
| \Psi_{{\rm gs},+}^{[\Omega_x = 0]} \rangle \}$.
The perturbation $\delta$ appears in the $H_{\rm eff}$ multiplied by $N$. This effect is the backbone of our quantum metrology protocol,
and it signals the amplification effect due to the spontaneous symmetry breaking that we will use to detect the $\delta$ field.

We note that the term $\Omega_{x}J_{x}$ couples the state $| \Psi_{{\rm gs},\mp}^{[\Omega_{x}=0]}\rangle$ to the next excited states
$| \Phi_{0,\mp(N/2-1)}\rangle$. In the superradiant phase this coupling perturbs the ground state multiplet, such that
\begin{equation}
|\Psi_{{\rm gs},\mp}^{[\Omega_x = 0]}\rangle\rightarrow
|\Psi_{{\rm gs},\mp}^{[\Omega_x = 0]}\rangle+e^{-\frac{2}{N}(\frac{g}{\omega})^{2}}
\frac{\Omega_{x}}{2\Omega_{x,\rm c}}\frac{\sqrt{N}}{1-1/N}|\Phi_{0,\mp(N/2-1)}\rangle,\label{pert}
\end{equation}
up to normalization factor. This perturbation, eventually gives a correction to the last term in (\ref{Heff}),
\begin{equation}
\frac{N \delta}{2} \sigma^z \to
\frac{N \delta}{2} \left( 1 - e^{-\frac{4}{N}
\left( \frac{g}{\omega}\right)^2}
\frac{\Omega_x^2}{2 \Omega^2_{x,{\rm c}}} \frac{1}{(1-1/N)^2}\right) \sigma^z,
\end{equation}
which can be neglected in the strongly coupled phase ($\Omega_x \ll \Omega_{x,{\rm c}}$).

Finally, we also note that the validity of $H_{\rm eff}$, together with the scaling given by Eq. (\ref{gap.1}), is an indirect proof of
the symmetry breaking of the parity symmetry anticipated by the expression (\ref{SSB}).

\section{Separation of time-scales for preparation and measurement}\label{PMS}
Our scheme relies on the adiabatic evolution of the system by considering a time-dependent transverse field $\Omega_x(t)$. Alternative versions of this scheme may consider the time variation of the coupling constant, $g$.
We assume that the system can be prepared in a linear superposition of low-energy states during an initial preparation stage (i),
which subsequently evolves quasi-adiabatically to perform a self-induced quantum many-body metrological stage (ii):

{\it (i) Preparation stage.-}
We consider an exponential decay
\begin{equation}
\Omega_x(t) = \Omega_x(0) e^{-t/\tau^{(1)}_{\rm ev}},
\end{equation}
with $\Omega_x(0) \gg \Omega_{x,{\rm c}}$, such that the system can be prepared initially in the ground state of the non-interacting phase, given by Eq. (\ref{psi0.1})
\begin{equation}
| \Psi(0) \rangle = | \Psi_{{\rm gs},-} ^{\rm [g = 0]} \rangle.
\end{equation}
The system evolves from $t=0$ up to $t = t_i$,
the latter being the initial time for the subsequent stage.
The transverse field varies up to $\Omega_{x}(t_i) = \Omega_{x,i}$, with $\Omega_{x,i} \ll \Omega_{x,{\rm c}}$, such that the system evolves into the strongly coupled regime.
Within the preparation stage the gap is bounded by
\begin{equation}
\Delta_i =\Delta_N \left(g, \omega, \Omega_{x,i} \right) .
\end{equation}
We impose full adiabaticity of the evolution of the system during the preparation stage,
\begin{equation}
1/\tau^{(1)}_{\rm ev} \ll \Delta_i.
\label{cond.1}
\end{equation}
Finally, we also need the condition,
\begin{equation}
\Delta_i \gg N \delta,
\label{cond.2}
\end{equation}
such that the system enters into the superradiant phase as an eigenstate of the $\Delta_N(g,\omega,\Omega_x)$ term in Hamiltonian
(\ref{Heff}). A crucial observation is that conditions
(\ref{cond.1}) and (\ref{cond.2}) imply that the preparation rate $1/\tau_{\rm ev}^{(1)}$ is not bounded by the parameter $\delta$. Thus  increasing the precision in measuring $\delta$ does not require increasingly longer $\tau_{\rm ev}^{(1)}$.

{\it (ii) Metrological stage.-}
Once the system is within the strongly coupled phase, we can use the two-level system approximation discussed in the previous section.
This is the part of the protocol where the measurement of $\delta$ is performed, and we require the quantum evolution to be sensitive to $N \delta$. Thus, for $t > t_i$, one can choose a second time scale for the evolution of the system, given by $\tau_{\rm ev}^{(2)}$,
\begin{equation}
\Omega_x(t) = \Omega_{x,i} e^{-(t - t_i) / \tau^{(2)}_{\rm ev}}.
\end{equation}
Note that within the strongly coupled phase the gap follows the scaling given by Eq. (\ref{gap.1}), such that
\begin{equation}
\Delta_N(t) = \Delta_N(g,\omega,\Omega_x(t)) = \Delta_i e^{- \gamma (t-t_i)},
\label{deltaN.ev}
\end{equation}
with $\gamma = N/\tau^{(2)}_{\rm ev}$.
The quantum metrological protocol will rely now on the quasi-adiabatic time evolution of the system, which is hold for
\begin{equation}
\gamma\ll\Delta^{\prime},\label{non-ad}
\end{equation}
where $\Delta^{\prime}$ is the energy splitting from $\vert \Psi_{+}^{[{\rm \Omega_{x}=0}]}\rangle$ to the next excited energy level. The condition (\ref{non-ad}) ensures that the non-adiabatic transitions to the other excited states are suppressed. In the strongly coupled phase we have $\Delta^{\prime}\approx \Omega_{x,\rm c}(1-1/N)$, which implies that the required condition reads $\gamma\ll \Omega_{x,\rm c}$, for large $N$.

Within the two-level approximation the state vector can be written as a superposition
\begin{equation}
\vert \Psi(t) \rangle
= c_{+}(t) \vert \Psi_{{\rm gs},+}^{[\Omega_x = 0]} \rangle
+
c_{-}(t)\vert \Psi_{{\rm gs},-}^{[\Omega_x = 0]} \rangle,
\label{superposition}
\end{equation}
where $c_{\pm}(t)$ are complex probability amplitudes. The condition $\Delta_i \gg N \delta$, ensures that the system is initially in an eigenstate of $\sigma^x$, with $c_+(t_i) = 1/\sqrt{2}$, and $c_-(t_i) = - 1/\sqrt{2}$.
The system evolves from $t = t_i$ up to a final time $t = t_f$, such that ends up in a phase
\begin{equation}
\Delta_N \left( g, \omega, \Omega_x(t_f) \right) =  \xi \ N \delta,
\end{equation}
with $\xi \ll 1$. In view of (\ref{deltaN.ev}), the latter condition can be re-written as
\begin{equation}
t_m = t_f - t_i = \frac{1}{\gamma} \log \left( \frac{\Delta_i}{\xi N \delta} \right).
\label{tm}
\end{equation}
Thus, up to logarithmic corrections, the measurement time, $t_m = t_f - t_i$, is directly governed by the rate $\gamma$.
\section{Quantum metrology protocol}\label{QMP}
In this section we focus on the description of the quasi-adiabatic evolution of the system during stage (ii) of the last section.
We have to solve the quantum evolution of a
two-level system with an exponentially decreasing transverse field, which turns out to be represented by the Demkov model with coupling $\Delta_{N}(t)=\Delta_i e^{-\gamma (t - t_i)}$
and detuning $N \delta$ \cite{Vitanov93jpb}.
Remarkably, the solution of the time-dependent Schr\"{o}dinger equation $\text{i}\frac{d}{dt}\vert\Psi\rangle= H_{\rm eff}\vert \Psi\rangle$
can be found exactly (see Appendix \ref{STSP}).

In the limit $xe^{-\gamma t_{m}}\ll 1$,
with $x=(\Delta_{i}/2\gamma)$, we obtain
\begin{equation}
\vert c_{+}(t_{f})\vert^{2}=\frac{1}{2}+\text{i}\frac{\pi}{4}\frac{x}{\cosh(\frac{\pi N\delta}{2\gamma})}
\{J_{\nu}(x)J_{-\nu}(x)-J_{\nu -1}(x)J_{1-\nu}(x)\},\label{solution}
\end{equation}
where $J_{\nu}(x)$ is a Bessel function of the first-kind \cite{Abramowitz} with $\nu=1/2-{\rm i}N\delta/2\gamma$.
For large $x\gg1$ we can use the asymptotic expansion $J_{\nu}(x)\sim \sqrt{2/\pi x}\{\cos(x-\frac{\pi\nu}{2}-\frac{\pi}{4})+O(x^{-1})\}$,
which yields for the $z$th component of the total angular momentum,
\begin{eqnarray}
&\langle J_{z}(t_{f})\rangle&=N\left(\vert c_{+}(t_{f})\vert^{2}-\frac{1}{2}\right)\nonumber\\
&&\approx-\frac{N}{2} \tanh \left(\frac{\pi N \delta}{2\gamma}\right)+O(x^{-1}).\label{Jz}
\label{mean.Jz}
\end{eqnarray}
The result represents the measured signal at $t_{f}\gg \gamma^{-1}$, as a function of $\delta$.
For vanishing perturbation field $\delta=0$
the final state is an equal superposition of the states (\ref{ground.state.1}), which yields $\langle J_{z}\rangle =0$.
However, for $\delta\neq 0$, the parity symmetry of $H_{\rm D}$ is broken and consequently of that the final probability amplitudes
$c_{\pm}(t_{f})$ are different, which allow us directly to estimate $\delta$ by measuring the collective spin population.
Depending on the ratio between typical values of $N \delta$ and $\gamma$ we have to distinguish the two following cases.

\subsection{Quasi-adiabatic protocol}
For $N\delta<\gamma$ the system evolution is a quasi-adiabatic in the sense that the dynamics is captured within the two-level subspace, but non-adiabatic effects within that subspace are used to estimate $\delta$.
Because the symmetry breaking term $H_{\delta}$ does not commute with the Dicke Hamiltonian $H_{\rm D}$ results in entangled superposition of the states (\ref{ground.state.1}) with probability amplitudes, depending the sign and magnitudes of $\delta$.
The measured signal at time $t_{f}$ is given by Eq. (\ref{Jz}) and the variance of the signal is
\begin{equation}
\langle\Delta^{2} J_{z}\rangle^{1/2}=\frac{N}{2\cosh\left(\frac{\pi N \delta}{2\gamma}\right)}.
\end{equation}
The uncertainty in measuring $\delta$ is given by
\begin{equation}
\bar{\delta}=\frac{\langle\Delta^{2} J_{z}\rangle^{1/2}}{\vert \partial \langle J_{z}\rangle/\partial \delta\vert}=\frac{2\gamma}{\pi N}\cosh\left(\frac{\pi N \delta}{2\gamma}\right),
\end{equation}
which is approximated with the Heisenberg-limited precision, $\bar{\delta}\approx 2\gamma/\pi N$.
\subsection{Full adiabatic protocol}
A different scheme can be devised by choosing a quantum evolution that is slower than typical values of $N \delta$. If $N \delta \gg \gamma$ the system evolution is dominated by the term $H_{\delta}$ in (\ref{H}), and we expect the system to follow adiabatically the ground state up to
$\vert \Psi_{{\rm gs},+}^{\Omega_x = 0} \rangle$
for
$\delta>0$ or $\vert \Psi_{{\rm gs},-}^{\Omega_x = 0} \rangle$ for $\delta<0$.
Thus, we expect the following approximation to hold,
\begin{equation}
\langle J_{z}(t_{f})\rangle\approx-\frac{N}{2} \frac{\delta}{|\delta|}\label{JzA}.
\end{equation}

We elaborate on this observation to devise a quantum metrological protocol that relies on a single-shot measurement of the spin-population to detect sign of $\delta$.
Let us assume that our initial knowledge of $\delta$ is given by a constant probability distribution within the interval
$[-\Delta_c, \Delta_c ]$
\begin{eqnarray}
P(\delta) = \frac{1}{2 \Delta_c}, \hspace{1cm} & & {\rm if} \ \ \
|\delta| \leq \Delta_c
\nonumber \\
P(\delta) = 0             , \hspace{1cm} & & {\rm if} \ \ \
|\delta| > \Delta_c.
\end{eqnarray}
Let us define the conditional probability
$P(\delta > \delta_c | S_z = - N/2 )$ as the probability that
$\delta > - \delta_c$ if we measure the value $- N/2$ of the observable $S_z$, with $\delta_c > 0$ and $\delta_c < \Delta_c$. In this way, $\delta_c$ is relates to the accuracy with which the adiabatic evolution allows us to measure the sign of the detuning $\delta$.
Similar definitions for the probabilities of values of $\delta$ are used below.
We can write
\begin{equation}
P(\delta > - \delta_c | S_z = -N/2) =
\int^\infty_{-\delta_c} P(\delta = \delta' | S_z = -\frac{N}{2}) d \delta'.
\end{equation}
The following expression can be obtained by means of Bayes' theorem,
\begin{equation}
P(\delta = \delta' | S_z = -\frac{N}{2}) = P(S_z = - \frac{N}{2} | \delta = \delta')
\frac{P(\delta')}{P(S_z = - \frac{N}{2})},
\end{equation}
Finally, taking the limit
$\Delta_c \ll \delta_c \ll \gamma/N$, and using Eq. (\ref{mean.Jz}), we find
\begin{equation}
P(\delta > - \delta_c | S_z = -N/2) =
1 -
\frac{\gamma}{2 \Delta_c \pi N} e^{- \frac{2 \pi \delta_c N}{\gamma}}.
\end{equation}
Note that this equation predicts that our quantum metrological protocol allows us to measure by a single-shot measurement the sign of $\delta$ with an error of $\gamma/N \approx 1/(t_m N)$ (up to logarithmic corrections), with $t_m$ the measurement time,
Eq. (\ref{tm}).
We also highlight that our method allows one to find a narrow spectral line even when the field is far-detuned. In contrast to the usual Ramsey spectroscopy, where such far-detuned field would not give any directional signature, due to the oscillation of the Ramsey signal \cite{Wineland94pra}.

Finally, we present some numerical results to check the validity of the two-state approximation used for our quantum metrological protocol. We compare the analytical result for $\langle J_{z}\rangle$
obtained by the Demkov model with the exact numerical solution of the time-dependent Schr\"{o}dinger equation with Hamiltonian (\ref{H}).
Figure \ref{fig3}a shows the measured signal as a function of $\delta$ for various $\gamma$.
In a quasi-adiabatic region, the signal is well approximated with
Eq. (\ref{Jz}),
while in the full adiabatic limit the signal tends to a step function, Eq. (\ref{JzA}).
In Fig. \ref{fig3}b we have checked the expression (\ref{Jz}) with the numerical exact result for various $N$. Finally, in Fig. \ref{fig4} we plot the measured signal as a function of $\gamma$ for various $N$. Remarkably, the exact solution follows Eq. (\ref{JzA}) for wide range of $\gamma$. In the limit $\gamma\gg N\delta$ the system dynamics become insensitive to $\delta$ in a sense that the signal $\langle J_{z}\rangle$ vanishes.

\section{Physical implementation with trapped ions}\label{PITI}
A linear crystal of trapped ions is an ideal system for the realization of our quantum metrology protocol.
Consider a chain of $N$ trapped ions with mass $M$ confined in a linear Paul trap along the $z$ axis. We assume that the effective spins are two internal states $\left\vert \uparrow \right\rangle$ and
$\left\vert \downarrow \right\rangle$ with frequency splitting $\omega_{0}$. Our protocol is intended to measure the detuning of laser with respect to $\omega_0$, for example to lock the frequency of the laser to the atomic internal transition. The interaction-free Hamiltonian describing the ion chain is given by
\begin{equation}
H_{0}=\sum_{i=1}^{N}\frac{\omega_{0}}{2}\sigma_{i}^{z}+\sum_{p=1}^{3N}\omega_{p}a_{p}^{\dag}a_{p},
\end{equation}
where $a_{p}$ and $a_{p}^{\dag}$ are the annihilation and creation operators of the $p$th vibration mode of the chain with corresponding frequency $\omega_{p}$.

\begin{figure}[h]
\includegraphics[width=0.35\textwidth]{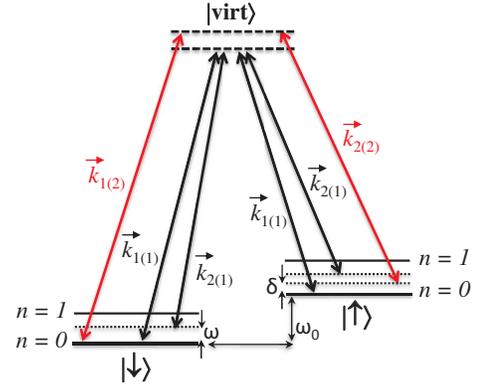}
\caption{The realization of the model (\ref{H}) based on a trapped ion setup. The linear ion crystal is uniformly addressed with two Raman laser beams with
wave-vector difference $\Delta \vec{k}_{1}=\vec{k}_{1(1)}-\vec{k}_{2(1)}$ pointing along the transverse $x$ direction. Spin-dependent force is created
by tuning the laser frequency close to the c.m. vibrational mode with detuning $\omega$. Additionally, the ion chain interact with pair of copropagating laser beams, i.e.,
$\Delta \vec{k}_{2}=\vec{k}_{1(2)}-\vec{k}_{2(2)}=0$ (no motional dependence) which drive the two-photon stimulated-Raman transition between
$\left\vert \uparrow\right\rangle$ and $\left\vert \downarrow\right\rangle$ spin states. $\delta$ is the laser detuning with respect to the frequency splitting $\omega_{0}$. }
\label{fig5}
\end{figure}

We consider that the ion chain is addressed collectively by means of two pairs of laser beams in a
Raman configuration as is shown in Fig. \ref{fig5}. We assume
that the first two non-copropagating laser beams have a wave-vector difference $\Delta \vec{k}_{1}$
along the transverse $x$ direction and laser frequency difference
$\Delta\omega_{1,\rm L}=\omega_{\rm c.m.}-\omega$, tuned near the center-of-mass (c.m.) vibrational mode $\omega_{\rm c.m.}$ with
detuning ($\omega_{\rm c.m.}\gg\omega$). Such a laser configuration generates a spin-dependent force, which provides a coupling between the effective spins and the c.m. mode.
The second pair of co-propagating lasers with frequency
difference $\Delta\omega_{2,\rm L}=\omega_{0}-\delta$ ($\omega_{0}\gg\delta$) induces a two-photon Raman transition between the spin states.
The Hamiltonian describing the laser-ion interaction becomes
\cite{SchneiderRPP,Nevado13epjst}
\begin{equation}
H_{\rm I}=\Omega\sum_{i=1}^{N}(e^{{\rm i}(\vert\Delta\vec{k}_{1}\vert x_{i}-\Delta\omega_{1,{\rm L}}t)}+{\rm h.c})\sigma_{i}^{z}+\Omega_{x}\sum_{i=1}^{N}(e^{-{\rm i}\Delta\omega_{2,{\rm L}}t}+{\rm h.c})\sigma_{i}^{x},
\label{H_I}
\end{equation}
where $\Omega$ and $\Omega_{x}$ are the respective interaction strengths.
Next, we transform
the trapped ion Hamiltonian in the rotating frame by means of
$U(t)=e^{-{\rm i}\{\frac{1}{2}(\omega_{0}-\delta)\sum_{i=1}^{N}\sigma_{i}^{z}+\sum_{p=1}^{3N}(\omega_{p}-\omega)a^{\dag}_{p}a_{p}\}}$,
and assume the Lamb-Dicke limit, which yields
\begin{equation}
H_{0}+H_{\rm I}\stackrel{U(t)}{\longrightarrow}H_{\rm D}(\omega,\Omega_{x},g)+H_{\delta}(\delta)+H^{\prime}(t),
\end{equation}
where $g=\eta \Omega$ is the spin-phonon coupling with $\eta=1/\sqrt{2 M\omega_{\rm c.m.}}$ being the Lamb-Dicke parameter ($\eta \ll 1$).
Here $H^{\prime}(t)$ describes
fast-rotating terms that can be neglected as long as $\vert \omega_{\rm c.m.}-\omega_{p\neq {\rm c.m.}}\vert \gg g,\omega$ and
$\vert \omega_{0}+\omega_{2, {\rm L}}\vert \gg \Omega_{x}$, respectively \cite{Nevado13epjst,Ivanov2013}.
The first condition ensures that within the motional rotating-wave approximation (r.w.a.) all vibration modes
can be neglected except the c.m. mode. A more detailed discussion on the conditions under which this approximation is
valid can be found in \cite{Ivanov2013}.
The second condition is the usual optical r.w.a. which ensures a pure $\sigma^x$ interaction.
Typical values in trapped ion systems could consist of a spin-phonon coupling $g=30$ kHz and effective boson frequency $\omega = 90$ kHz. This choice corresponds to the results
presented in Fig. \ref{fig3}.

For $N=10$ and $\gamma=0.1 g$ we estimate that the initial state is transformed into the final state (\ref{superposition})
with probability amplitudes given by Eq. (\ref{solution}) approximately for $23$ ms,
which is comparable with the experimentally measured coherence time in typical trapped ion setups \cite{Benhelm08,Monz11}. Further increasing of the
coherence time could be achieved either by using a magnetic insensitive clock states \cite{Aolitai_2011} or decoherence-free qubit states \cite{Ivanov10}. Finally, the collective spin population can be measured by laser induced fluorescence, which is imaged on a CCD camera.

It is essential for our protocol to rely on a $\sigma^z$ spin-phonon coupling that yields the parity symmetric Hamiltonian $H_{\rm D}$.
Additionally, ac-Stark shifts have to be reduced to the point that they are neglected compared to the sensitivity in the
estimation of the detuning $\delta$, but fluctuations in the laser intensity will limit the cancellation of those terms.
A particularly well-suited configuration to achieve both $\sigma^z$ coupling and cancellation of ac-Stark shifts is
provided by ions trapped in Penning traps, see for example the scheme shown in \cite{Britton12nat},
where a $\sigma^z$ interaction together with the cancellation of ac-Stark shifts is achieved with a
proper configuration of laser polarizations. We also highlight that a very similar protocol to the
one introduced here could be used for estimation of a displacement term $\delta \sum_j( a_j + a_j^\dagger )$,
the latter playing the role of a symmetry breaking perturbation. This could allow to devise adiabatic quantum metrological
schemes for ultra-sensitive detection of forces \cite{Biercuk10natnano}.

\section{Conclusions and Outlook}\label{Conclusion}
We have studied the process of symmetry breaking of a discrete symmetry due to the presence of small perturbation field in a system described by the Dicke Hamiltonian.
We have shown that quasi-adiabatic evolution in this system induces a quantum metrology protocol, which is Heisenberg limited. Our many-body Ramsey spectroscopy protocol can be implemented with linear ion crystal, where the symmetry breaking field is controlled by the laser detuning to the respective
qubit transition. The realization of the proposed quantum metrology protocol is not restricted
only to trapped ions but could be implemented with other experimental setups such as cavity \cite{Dimer07} or circuit QED \cite{Nataf10} systems.

We highlight a few advantages of our idea with respect to current approaches to quantum metrology: (i) Our method does not require quantum gates, since it is induced by always-on interactions. (ii) In principle, our work does not rely on effective spin-spin interactions mediated by auxiliary photonic or bosonic fields. On the contrary, our adiabatic process may also work in a regime in which $g \geq \omega$, such that the final state is not a pure state of qubits, but an entangled spin-boson state instead. (iii) Since our method mainly relies on symmetry considerations, it should be robust with respect to perturbations to $H_{\rm D}$ that respect the parity symmetry. (iv) We note that our method allows us to get information about $\delta$ with a single-shot measurement in the full adiabatic scheme.

We also remark that the scheme presented here share some of the limitations as standard protocols with quantum metrology with NOON states \cite{Huelga97prl}. In particular, our method would not imply any advantage if the measurement time is limited by decoherence.
Also, an important limitation of our scheme is the fact the spin-boson interactions have to be fully parity symmetric, being any deviation from that symmetry a potential source of error in the achieved accuracy.

We finish with an Outlook of possible research directions motivated by this work. We have presented a very specific study relying on a model belonging to the long-range Ising universality class. It would be very interesting to explore scalings related to similar quantum metrology protocols with different universality classes and symmetries, like those that can be simulated with trapped ions, for example
\cite{Porras04aprl,Ivanov09pra,Bermudez09pra}. Also, one could study quantum dissipative phase transitions
\cite{Horstmann13pra,Muller13AAMOP} in addition to the evolution of closed quantum systems presented here. Finally, although we have presented an example with trapped ions and frequency estimation, one could also think of applications to measure forces or magnetic fields, for example.

\acknowledgments{
This work has been supported by the Bulgarian NSF grant DMU-03/107, Spanish projects QUITEMAD S2009-ESP-1594, RyC Contract No. Y200200074, and the European COST Action MP IOTA 1001. We thank K. Singer,
S. Dawkins and P.O. Schmidt for useful discussions.}

\appendix
\section{Calculation of the gap for the Dicke model}\label{EG}
For $\Omega_{x}=0$, the energy spectrum of the Dicke Hamiltonian $H_{\rm D}$ can be analytically carried out by a simple canonical transformation,
namely
\begin{equation}
\tilde{H}=D^{\dag}(\alpha)H_{\rm D}D(\alpha)=\omega a^{\dag}a-4\left(\frac{g^{2}}{N\omega}\right)J_{z}^{2},
\end{equation}
with displacement operator defined by
\begin{equation}
D(\alpha)=\exp{(\alpha a^{\dag}-\alpha^{\dag}a)},\quad \alpha=-2\left(\frac{g}{\omega}\right)\frac{J_{z}}{\sqrt{N}}.
\end{equation}
The eigenvalues and eigenvectors of $H_{\rm D}$ are
\begin{equation}
E_{n,m}=n\omega-\frac{g^{2}N}{\omega}\left(\frac{m}{N/2}\right)^{2}\label{Energy}
\end{equation}
and
\begin{equation}
\vert \Phi_{n,m}\rangle = D(\alpha)\vert j,m \rangle\vert n \rangle_{\rm b}\label{Vectors}.
\end{equation}
Here, the Hilbert space of the system consists of the state $\{\vert j,m\rangle\otimes\vert n\rangle_{\rm b}$\}, where $\vert j,m\rangle$ ($m=-j,\ldots,j$) are the Dicke states, $\vec{J}^{2}\vert j,m\rangle=j(j+1)\vert j,m\rangle$, $ J_{z}\vert j,m\rangle=m\vert j,m\rangle$ and $\vert n\rangle_{\rm b}$ is the Fock state of the bosonic field mode with occupation number $n$. The energy spectrum of $H_{\rm D}$ is a double degenerate with ground state energy $E_{{\rm gs},\mp}=-N(g^{2}/\omega)$ and corresponding eigenvectors
\begin{equation}
\vert \Psi_{{\rm gs},\pm }^{[\Omega_x = 0]} \rangle=D(\alpha_{\pm }) \vert \frac{N}{2},\pm \frac{N}{2} \rangle\vert 0 \rangle_{\rm b}.
\end{equation}
with $\alpha_{\pm}=\mp \sqrt{N}(g/\omega)$.

The term $\Omega_{x}J_{x}$ split the degeneracy of the energy spectrum and thus creates an effective coupling between the states $\vert \Phi_{n,\pm m}\rangle$. Assuming $\Omega_{x}\ll g,\omega$, the effect of the latter can be treated by perturbation theory. The splitting between the two lowest energy eigenstates is given by
\begin{widetext}
\begin{eqnarray}
\Delta_N(g,\omega,\Omega_x)=2\Omega_{x}^{N}\left|\sum_{n_{1},n_{2}\ldots n_{N-1}}\frac{\langle \Psi_{{\rm gs},-}\vert J_{x}\vert \Phi_{n_{1},-\frac{N}{2}+1}\rangle \langle \Phi_{n_{1},-\frac{N}{2}+1}\vert J_{x}\vert \Phi_{n_{2},-\frac{N}{2}+2} \rangle \ldots \langle \Phi_{n_{N-1},\frac{N}{2}-1}\vert J_{x}\vert \Psi_{{\rm gs},+} \rangle}{(E_{{\rm gs},-}-E_{n_{1},-\frac{N}{2}+1})(E_{{\rm gs},-}-E_{n_{2},-\frac{N}{2}+2})\ldots (E_{{\rm gs},-}-E_{n_{N-1},\frac{N}{2}-1})}\right|.\label{Gap}
\end{eqnarray}
\end{widetext}
(We assume $\Omega_x = 0$ for state-vectors and energies in the latter expression and in the rest of the Appendix). For weak coupling $g\ll\omega$ the bosonic mode is only virtually excited in a sense that it only transmits the effective spin-spin interaction. This allows to simplify the expression Eq. (\ref{Gap}) as follows
\begin{widetext}
\begin{eqnarray}
\Delta_N(g,\omega,\Omega_x)=2\Omega_{x}^{N}\left|\frac{\langle \Psi_{{\rm gs},-}\vert J_{x}\vert \Phi_{0,-\frac{N}{2}+1}\rangle \langle \Phi_{{\rm gs},-\frac{N}{2}+1}\vert J_{x}\vert \Phi_{0,-\frac{N}{2}+2} \rangle \ldots \langle \Phi_{0,\frac{N}{2}-1}\vert J_{x}\vert \Psi_{{\rm gs},+} \rangle}{(E_{{\rm gs},-}-E_{0,-\frac{N}{2}+1})(E_{{\rm gs},-}-E_{0,-\frac{N}{2}+2})\ldots (E_{{\rm gs},-}-E_{0,\frac{N}{2}-1})}\right|.\label{AGap_WC1}
\end{eqnarray}
\end{widetext}
Using, Eqs. (\ref{Energy}) and (\ref{Vectors}) the energy gap (\ref{AGap_WC1}) reads
\begin{equation}
\Delta_N(g,\omega,\Omega_x)=2e^{-2\left(\frac{g}{\omega}\right)^{2}}\frac{N^{N+1}}{2^{N}\Gamma (1+N)}\Omega_{x}\left(\frac{\Omega_{x}}{\Omega_{x,\rm c}}\right)^{N-1}.
\end{equation}
The asymptotic behavior of $\Delta_N(g,\omega,\Omega_x)$ for large $N$ can be derived by using Stirling's formula $\Gamma(1+z)\sim \sqrt{2\pi z}z^{z}e^{-z}$, which yield
\begin{equation}
\frac{\Delta_N(g,\omega,\Omega_x)}{\Omega_{x}}\sim  \sqrt{\frac{2}{\pi}}e^{-2\left(\frac{g}{\omega}\right)^{2}}\left(\frac{\Omega_{x,\rm c}}{\Omega_{x}}\right)\sqrt{N}e^{-N\{\ln\left(\frac{2\Omega_{x,\rm c}}{\Omega_{x}}\right)-1\}}.
\end{equation}
\section{Exact solution of the Demkov model}\label{STSP}
The two-state problem consists of the following system of differential equations:
\begin{eqnarray}
&&{\rm i}\dot{c}_{+}(t)=\frac{N\delta}{2} c_{+}(t)+\frac{\Delta_{N}(t)}{2}c_{-}(t),\nonumber \\
&&{\rm i}\dot{c}_{-}(t)=-\frac{N\delta}{2} c_{-}(t)+\frac{\Delta_{N}(t)}{2}c_{+}(t).\nonumber \\
\label{system}
\end{eqnarray}
Here $N\delta$ is a constant, while the effective coupling depends on time $\Delta_{N}(t)=\Delta_{i}e^{-\gamma t}$,
which reduces the two state problem to the Demkov model. We seek the solution of Eq. (\ref{system})
assuming the initial conditions $c_{+}(0)=1/\sqrt{2}$ and $c_{-}(0)=-1/\sqrt{2}$.
The latter correspond to the ground state of Hamiltonian (\ref{Heff}) in the limit $\Delta_{i}\gg N\delta$.

The system (\ref{system}) can be decoupled by differentiating with respect to $t$, which yield
\begin{eqnarray}
&&\ddot{c}_{+}(t)+\gamma \dot{c}_{+}(t)+\{\left(\frac{N\delta}{2}\right)^{2}+\text{i}\frac{\gamma N\delta}{2}+\frac{\Delta_{i}^{2}}{4}e^{-2\gamma t}\}c_{+}(t)=0,\nonumber \\
&&\ddot{c}_{-}(t)+\gamma \dot{c}_{-}(t)+\{\left(\frac{N\delta}{2}\right)^{2}-\text{i}\frac{\gamma N\delta}{2}+\frac{\Delta_{i}^{2}}{4}e^{-2\gamma t}\}c_{-}(t)=0.\nonumber \\
\label{second_order}
\end{eqnarray}
Next, we introduce a dimensionless time $z=x e^{-\gamma t}$ with $x=\Delta_{i}/2\gamma$, which transforms the set of equations (\ref{second_order}) to
\begin{eqnarray}
&&z^{2}\ddot{c}_{+}(z)+\{z^{2}+\left(\frac{N\delta}{2\gamma}\right)^{2}+{\rm i}\left(\frac{N\delta}{2\gamma}\right)\}{c}_{+}(z)=0,\nonumber \\
&&z^{2}\ddot{c}_{-}(z)+\{z^{2}+\left(\frac{N\delta}{2\gamma}\right)^{2}-{\rm i}\left(\frac{N\delta}{2\gamma}\right)\}{c}_{-}(z)=0.\nonumber \\
\end{eqnarray}
The solution can be written as \cite{Abramowitz}
\begin{eqnarray}
&&c_{+}(z)=\sqrt{z}\{a_{1}J_{\nu}(z)+a_{2}J_{-\nu}(z)\},\nonumber \\
&&c_{-}(z)=\sqrt{z}\{b_{1}J_{1-\nu}(z)+b_{2}J_{\nu-1}(z)\}.\nonumber \\
\end{eqnarray}
Here $J_{\nu}(z)$ is a Bessel function of the first kind \cite{Abramowitz} with $\nu=1/2-{\rm i}N\delta/2\gamma$. The constants $a_{1,2}$ and $b_{1,2}$ can be determined by the initial conditions at $t=0$. We find
\begin{equation}
c_{+}(t)=\frac{\pi}{2\sqrt{2}}\frac{e^{-\frac{\gamma t}{2}}x}{\cosh \left(\frac{\pi N \delta}{2\gamma}\right)}\{a_{1}J_{\nu}(x e^{-\gamma t})+a_{2}J_{-\nu}(x e^{-\gamma t}\},
\end{equation}
with
\begin{eqnarray}
&&a_{1}=J_{1-\nu}(x)-{\rm i}J_{-\nu}(x),\nonumber \\
&&a_{2}=J_{\nu -1}(x)+{\rm i}J_{\nu}(x).\nonumber \\
\end{eqnarray}
and $\vert c_{-}(t)\vert ^{2}=1-\vert c_{+}(t)\vert ^{2}$.

In the limit $x e^{-\gamma t_{f}}\ll 1$ one can derive an asymptotic form of the probability $\vert c_{+}(t_{f})\vert ^{2}$ by using $J_{\nu}(z)\sim \frac{1}{\Gamma(1+\nu)}(z/2)^{\nu}$, which yield
\begin{equation}
\vert c_{+}(t_{f})\vert^{2}=\frac{1}{2}+\text{i}\frac{\pi}{4}\frac{x}{\cosh(\frac{\pi N\delta}{2\gamma})}
\{J_{\nu}(x)J_{-\nu}(x)-J_{\nu -1}(x)J_{1-\nu}(x)\}.
\end{equation}
In the above expression we have used the identities $\Gamma(\nu)\Gamma(1-\nu)=\pi/\sin(\pi\nu)$ and $J_{\nu-1}(x)J_{-\nu}(x)+J_{\nu}(x)J_{1-\nu}(x)=2\sin(\pi\nu)/\pi x$, respectively. Finally, for $x\gg 1$ and $x\gg \vert \nu^{2}-1/4\vert$ the Bessel function has the asymptotic form $J_{\nu}(x)\sim \sqrt{2}{\pi x}\cos(x-\frac{\pi \nu}{2}-\frac{\pi}{4})$, which gives
\begin{equation}
\vert c_{+}(t_{f})\vert^{2}=\frac{1}{2}-\frac{1}{2}\tanh\left(\frac{\pi N\delta}{2\gamma}\right).
\end{equation}

\bibliography{qm.bbl}




\end{document}